\documentclass{PoS}

\title{Hybrid morphology radio sources - follow-up VLBA observations}

\ShortTitle{HYMORS - follow-up VLBA observations}
\author{\speaker{Maciej Ceg\l{}owski}%
         \\
        Torun Centre for Astronomy, Gagarina 11, 87-100 Torun, Poland\\
        E-mail: \email{ceglowski@astri.uni.torun.pl}}
\author{Marcin Gawro\'nski\\
        Torun Centre for Astronomy, Gagarina 11, 87-100 Torun, Poland\\
        E-mail: \email{motylek@astro.uni.torun.pl}}
\author{Magdalena Kunert-Bajraszewska\\
        Torun Centre for Astronomy, Gagarina 11, 87-100 Torun, Poland\\
        E-mail: \email{magda@astro.uni.torun.pl}}

\abstract{Hybrid sources that present FR I - like jet on the one side of 
the radio core and FR II - like on the other are rare class of objects
that may posses key to understanding the origin of FR division. We presents
information connected with the new high resolution VLBA follow-up observations of 5 recently discovered
hybrid sources. We believe that sources which exhibit
two different morphologies at the opposite side of the radio core
are FR II type objects evolving in nonuniform high-density environment. 
}

\FullConference{12th European VLBI Network Symposium and Users Meeting\\
		7-10 October 2014\\
		Cagliari, Italy}

\begin{document}

\section{Introduction}

Large scale  radio sources may evolve into two different morphological 
classes: Fanaroff-Riley type I and type II sources (FRIs and FRIIs) \cite{fr}. 
FR II class is characterize by bright hot-spots evident at the end of jets
which are situated at opposite side of the radio core. Radio jets are usually not so prominent
and sometimes may not even be visible in so called FR II type objects.  
FRIIs resemble quite homogeneous radio structure.
On the contrary,  FR Is - shows much more diversity \cite{parma02}. 
Nearly half of those objects displays a mix of narrow-angle or wide-angle tailed sources (NATs and WATs). 
FR I sources usually posses prominent edge-darkening twin jets. 
What is more, FRIs do not show hotspots at the outer parts of the
radio structure.

In addition to this, statistically, FR IIs  
 displays higher radio luminositys. Fanaroff \& Riley
 in 1974 introduced threshold in order to distinguish
 brighter FR IIs from FR Is, $L_{178 MHz} \sim$
$10^{25.5}$ W~H$z^{-1}$ $sr^{-1}$.  Wall et al. (1980) \cite{wall} also 
states that those classes undergo different cosmological evolutions.

Existence of two distinct morphological classes has been widely confirmed for more than 40 years,
yet its origin still requires equivocal explanation.  Some authors believe that it 
is connected with transition of the intrinsically supersonic jet to transonic/subsonic flow,
which is being decelerated due to entrainment of thermal plasma in the innermost region of the 
host galaxy \cite{komi94,bow,bick,kia}. On the other hand, 
different morphology might be connected with the central engine \cite{mei,zib} or 
the jet itself \cite{rey, ghi}. The most recent scenario 
tries to connect FR\,I/FR\,II dichotomy to the interaction between radio jet and 
the intergalactic medium \cite{gopkri91,gopkri96}. This in turn may lead to 
disruption of the collimated jet. 
Interestingly, Gopal-Krishna \& Witta (2000) \cite{gopwii} found  a rare group of objects 
that indeed exhibits two different morphological classes of the jets, and they named them HYMORS ({\bf HY}brid {\bf MO}rphology 
{\bf R}adio {\bf S}ource). They suggest that the HYMORS class may be essential for 
understanding the FR\,I/FR\,II dichotomy. 
It is worth pointing out that existence of HYMORS speaks in favor of a hypothesis 
in which the intergalactic medium and jet power play a dominant role in shaping radio morphology.

Gawro\'nski et al. (2006) \cite{gawron06} discovered 5 new HYMORS sources. Their results strongly support existence of 
two different kinds of jets in HYMORS. It seems that indeed 
 a significant over-density of cold gas is present and might by essential in shaping
 large scale morphology. In recent survey  we  describe a follow-up observations \cite{ceglowski13} of the cores of the hybrid
sources discovered by Gawro\'nski et al. (2006).

Main motivation was to answer the question if the radio jet prefers orientation towards the FR I or FR II -like morphology at milli-arcsecond scales.

\section{Observations}
 Using the VLA Faint Images of the Radio Sky at Twenty-centimeters (FIRST) survey \cite{white97}
Gawro\'nski et al. (2006) discovered 5 HYMORS sources.  Main motivation of this work was to thoroughly investigate radio morphology 
identify in the FIRST maps. 
 
Based on more sensitive observations at 4.9 GHz from VLA large-scale hybrid structures was confirmed. 
Our recent survey  focused on the proximity of the central engine \cite{ceglowski13}.
We performed follow-up, new high resolution radio observations at the 1.7\,GHz and 5\,GHz using VLBA (26 \& 27 May 2007).
Each target source, together with its associated phase reference source, was observed for $\sim$3\,hrs 
per frequency in phase-referencing mode. The whole data reduction process was carried out using standard NRAO AIPS
procedures. 


\begin{figure*}[t]
\centering
\includegraphics[width=\textwidth]{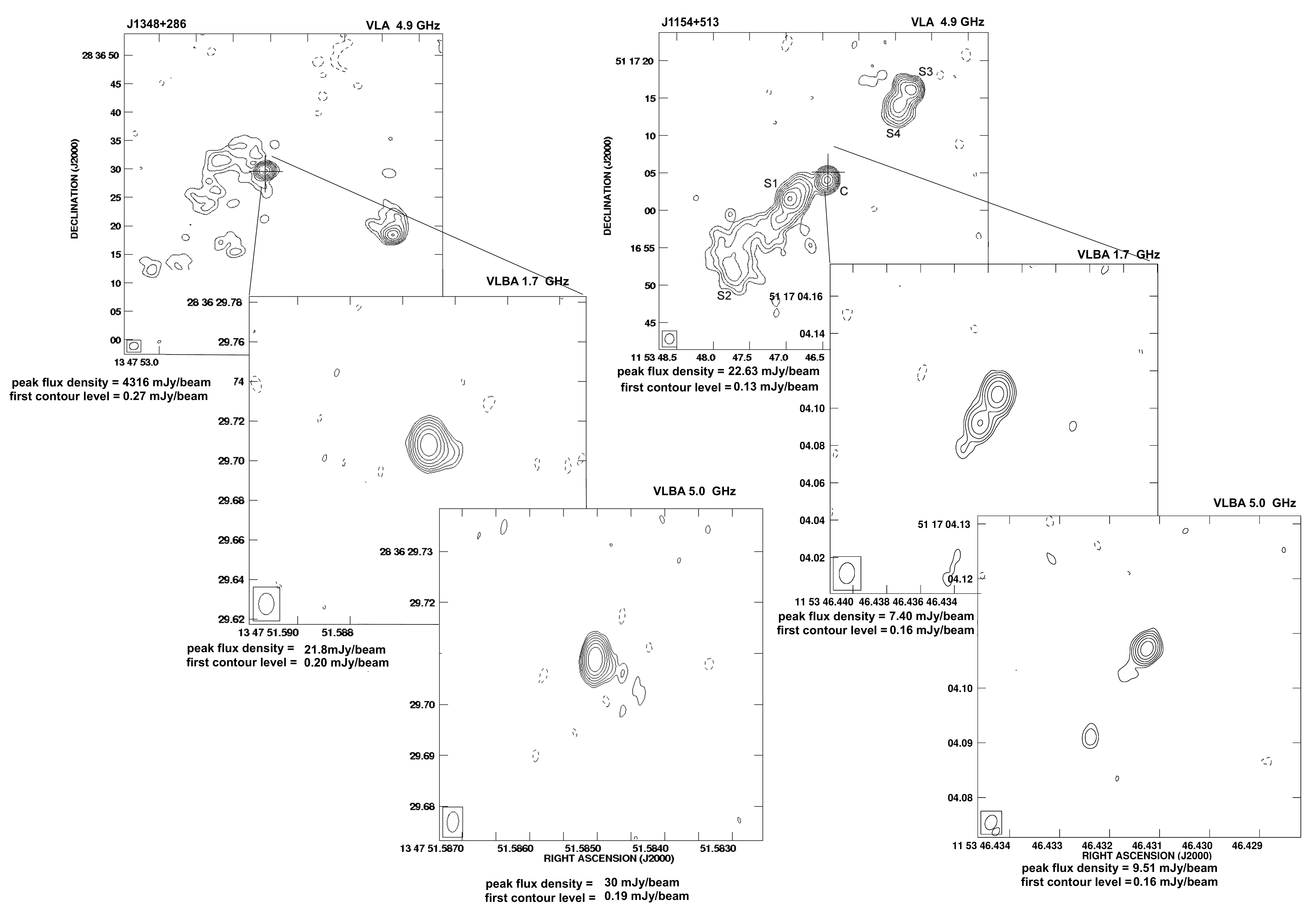} 

\caption{
HYMORS sources J1154+513 \& J1348+286 with zoomed-in radio core image. 
Upper radio map represents arcsecond scale taken from Gawro\'nski et al. (2006) \cite{gawron06}. Lower maps exhibits 
central engine proximity at VLBA L-band and C-band. Contours increase  
by a factor 2. The first contour level corresponds to $\approx  
3\sigma$. Position of an optical counterpart - indicated by a cross - is taken from the SDSS DR7.
 }   
\label{hybrids_1}
\end{figure*}


\section{Conclusions}

Interestingly, there are very few articles dedicated precisely to hybrids sources.
What is more, they usually present archival data. Hence, there is no complete morphological 
information about any of the previous studied samples of HYMORS.
As if this was not enough, in some cases the classification as hybrid objects is highly uncertain.

The new sample consisting of 5 hybrid objects, for the
first time, posses full information about their morphology, 
on both VLA and VLBA scales as well as luminosities, and beaming estimations.
Based on the radio core power at 5 GHz and radio power at 408 MHz
using approach proposed by Giovannini G. et al. (2001) \cite{gio01}
we estimated the Doppler factor.  
HYMORS sources presented in our recent survey are  unbeamed objects with viewing angles more than {\bf $21^{\rm o}$} for the 
inner parsec scale structures. What is more, we believe that VLBA observations do not suggest there is a preferable spatial orientation in hybrid objects. 
Two out of five sources revealed milliarcsecond core-jet structure and another two probable weak jets.
It seems that very weak parsec-scale radio jets could be present in all objects.
  Moreover, the estimated luminosity of observed hybrids sources
is significantly higher than the traditional FR\,I/FR\,II break luminosity indicating they have
radio powers similar to FR\,IIs.

Based on the performed analysis of the sample of new hybrid objects we
suggested that HYMORS are FR\,IIs evolving in a heterogeneous environment


\begin{thebibliography}{99}

\bibitem[1]{bick} Bicknell, G.~V., 1995, APJS, 101, 29 


\bibitem[2]{bow} Bowman, M., Leahy, 
J.~P., \& Komissarov, S.~S., 1996, MNRAS, 279, 899

\bibitem[3]{ceglowski13} Ceg{\l}owski, M., Gawro{\'n}ski, M.~P., \& Kunert-Bajraszewska, M.\ 2013, AAP, 557, AA75 


\bibitem[4]{fr} Fanaroff, B.~L., \& Riley, J.~M., 1974, MNRAS, 167, 31P

\bibitem[5]{ghi} Ghisellini G., Celotti A., 2001,
A\&A, 379, L1

 \bibitem[6]{gawron06} Gawro{\'n}ski, M.~P., Marecki, A., Kunert-Bajraszewska, M., \& Kus,  A.~J., 2006, AAP, 447, 63 

 \bibitem[7]{gopkri91} Gopal-Krishna, 1991, AAP, 248, 415

 \bibitem[8]{gopkri96} Gopal-Krishna, Wiita, P.~J., \& Hooda, J.~S.\ 1996, AAP, 316, L13

 \bibitem[9]{gopwii} Gopal-Krishna, \& Wiita, P.~J., 2000, AAP, 363, 507

\bibitem[10]{gio01} Giovannini G., Cotton W.~D., Feretti L. , Lara L., Venturi T.,
2001, APJ, 552, 508

\bibitem[11]{kia} Kaiser, C.~R., \& Alexander, P., 1997, MNRAS, 286, 215 



\bibitem[12]{komi94} Komissarov, S.~S., 1994, MNRAS, 269, 394 

\bibitem[13]{mei} Meier, D.~L., Edgington, 
S., Godon, P., Payne, D.~G., \& Lind, K.~R.\ 1997, {\emph Nature}, 388, 350

\bibitem[14]{parma02} Parma, P., Murgia, M., de 
Ruiter, H.~R., \& Fanti, R.\ 2002, NAR, 46, 313

\bibitem[15]{rey} Reynolds, C.~S., Di 
Matteo, T., Fabian, A.~C., Hwang, U., 
\& Canizares, C.~R.\ 1996, MNRAS, 283, L111 

\bibitem[16]{wall} Wall, J.~V., Pearson, 
T.~J., \& Longair, M.~S.\ 1980,MNRAS , 193, 683

\bibitem[17]{white97} White, R.~L., Becker, 
R.~H., Helfand, D.~J., \& Gregg, M.~D.,1997, ,ApJ, 475, 479

\bibitem[18]{zib} Zirbel, E.~L., \& Baum, S.~A.\ 1995, ApJ, 448, 521 

\end{thebibliography}
\end{document}